\documentclass[aps,pra,twocolumn]{revtex4}
\usepackage{graphicx}

\begin{document}
\draft
\preprint{Version \today}

\title{Siegert pseudostate perturbation theory: one- and two-threshold cases}
\author{Koudai Toyota,  Toru Morishita, and Shinichi Watanabe}
\affiliation{Department of Applied Physics and Chemistry,
The University of Electro-Communications,\\
1-5-1 Chofu-ga-oka, Chofu-shi, Tokyo 182-8585, Japan}

\begin{abstract}
Perturbation theory for the Siegert pseudostates (SPS)
[Phys.~Rev.~A{\bf 58}, 2077 (1998) and Phys.~Rev.~A{\bf 67}, 032714 (2003)]
is studied for the case of two energetically separated
thresholds.
The perturbation formulas for the one-threshold case are derived
as a limiting case whereby we reconstruct
More's theory for the decaying states [Phys.~Rev.~A{\bf 3},~1217~(1971)]
and amend an error.
The perturbation formulas for the two-threshold case have additional 
terms due to the non-standard orthogonality relationship of the 
Siegert Pseudostates.
We apply the theory to a 2-channel model problem, and 
find the rate of convergence of the perturbation expansion should be
examined with the aide of the variance $D= ||E-\sum_{n}\lambda^n E^{(n)}||$ 
instead of the real and imaginary parts of the perturbation energy individually.
\end{abstract}

\maketitle
\section{INTRODUCTION}
Resonances occur in a variety of fields of physical sciences. 
Despite their diversity, they are characterized by two parameters, the
resonance energy position and width, apart from the coupling with the
background continuum represented by the Fano profile\cite{fano-rau}.
A great deal of discussions have been given to the interpretation of
resonance phenomena\cite{fano-rau}.
The most
familiar parameterization of the resonances is condensed into the dispersion
formula due to Breit and Wigner.
 Back in 1939,
Siegert \cite{Siegert} developed a
compact mathematical viewpoint for characterizing resonances as singular
points of the dispersion relation.
His idea requires the solution
of the Schr\"odinger equation subject to the outgoing wave boundary condition,
\[
\left.\left(\frac{d}{dr}-ik\right)\phi(r)\right|_{r=a}=0,
\]
where $a$ is the radius beyond which the potential energy is negligible.
The solution $\phi(r)$ is called the Siegert state~(SS) and it behaves like 
${\rm e}^{i k r}$ near $r=a$ and beyond.
This boundary condition destroys the hermiticity of the Hamiltonian, thus
entailing complex-valued eigenenergies, {\it i.e.},
\[
 E=\frac{k^2}{2}=E_{\rm res}-i\frac{\Gamma}{2}
\]
This is a most direct representation of both the resonance position and
width.
This mathematically appealing representation had been implemented with
tedious iterations due to lack of suitable computational techniques until
Tolstikhin {\it et al} \cite{oleg} made a breakthrough
by introducing Siegert pseudostates (SPS) for the one-threshold case. Their
idea incorporates the boundary condition into the Schr\"odinger equation
so that the dispersion relation is obtained by a single diagonalization of
the Hamiltonian matrix. Previous applications of SPS to resonances in three-body Coulomb
problems indicate that it is not only a valid procedure but also a new
perspective for the SPS representation of resonances and decay
processes \cite{olegclmb}.
Another immediate application of the SPS theory is to the
time-dependent problem\cite{time-dept,CHG} where the reflection off the exterior
boundary incurs numerical instability. Tanabe and Watanabe~\cite{tana-wata}
succeeded in describing the reflectionless time propagation based on the
Siegert pseudostates. Indeed, applied to the half-cycle optical pulses, the 
Siegert boundary condition indeed was seen to eliminate the outgoing wave component
perfectly. 

Recently, Sitnikov and Tolstikhin \cite{sit-tol,toyota} stretched the scope of the SPS
theory by enabling the treatment of the two-threshold problem. Despite
such progress, 
there remains in the theory of SPS a chapter still
incompletely worked out. This is the Siegert perturbation theory. 
A pioneering work on this subject is due to More\cite{more1,more2} who
extended the Siegert state theory specifically to handle the decaying state.
The main purpose of this paper is to complete the Siegert perturbation
theory from the recently developed SPS viewpoint for both one- and
two-threshold cases. Particularly, in the one-threshold case, 
we are able to reconstruct More's theory for decaying states~\cite{more1}
in terms of SPS but with an unexpected amendment to his theory.
The SPS perturbation theory~(SPSPT) is by no means straightforward owing to the
non-standard orthonormality of the eigenfunctions. This constraint also
serves to fix the phase of the perturbed wavefunction, a feature which
is absent from the standard perturbation theory. It is hoped this paper
serves to expose such noteworthy features of the SPSPT.
 
This paper is thus constructed as follows.
In Section~\ref{sect:review}, we review some basic ideas about the
SPS as needed for an elementary presentation of the
perturbation theory. Section~\ref{sect:PT} gives the details of the
SPSPT for both one- and two-threshold cases. 
And Section~\ref{sect:PT} deals with a specific mathematical model as
an example of the SPSPT. Atomic units are used throughout.
\section{THE SIEGERT PSEUDOSTATES}
\label{sect:review}
Since the two-threshold SPS theory contains
the one-threshold case in itself, we review the two-threshold case only,
leaving the one-threshold case as
 the limit where the two-thresholds become degenerate~\cite{sit-tol}. 
\subsection{Mathematical Settings}
Suppose first that there are as many as $q$ independent channels.
The Schr\"odinger equation reads
\begin{equation}
\left[-\frac{1}{2}\frac{d^2}{dr^2}+V(r)-E\right]\phi(r)=0,
\label{eq:2ch}
\end{equation}
where
\begin{eqnarray*}
V(r)&=&
\left(\begin{array}{cccc}
V_1 &V_{12} & \cdots &V_{1q}\\
V_{12} &V_2 & \cdots &V_{2q}  \\
\vdots & \vdots & \ddots&\vdots\\
V_{1q} &V_{2q}& \cdots & V_q\\
\end{array}
\right),\\
\phi(r)&=&\left(
\begin{array}{c}
\phi_{1} \\
\phi_{2} \\
\vdots\\
\phi_{q} 
\end{array}
\right)
\end{eqnarray*}
and $V_{i}$ pertains to the potential energy of channel $i$, and
$V_{ij}$ represents the interchannel
coupling between channels $i$ and $j$. We consider the situation where
there are only two energetically distinct thresholds so that we
separate $V_{i}$  into two groups. A first group contains channels
$1,\ldots,q_{1}$ and they 
converge to $v_{1}$ as $r \to a$ while  
the other group contains channels $q_{1}+1,\ldots,q$ and they
converge to $v_{2}$, that is
\[
\lim_{r\rightarrow a}V(r)
={\rm diag}[
\overbrace{v_1,\cdots,v_1}
	^{1,\cdots,q_{1}},
\overbrace{v_2,\cdots,v_2}
	^{q_{1}+1,\ldots,q}
],
\]
where $v_1$  and $v_2$ are the two constants representing the threshold
energies.
This allows us to use the {\it 2-channel} SPS scheme even in the
presence of {\it more than two channels}.
The two channel momenta are $k_{1}=\sqrt{2(E-v_{1})}$ and
$k_{2}=\sqrt{2(E-v_{2})}$.
The boundary conditions are thus
\[
\phi_{i}(0)=0
\]
at $r=0$ and
\[
\left. \left(\frac{d}{dr}-ik_{j}\right)\phi_{i}\right|_{r=a}=0
\]
at $r=a$ where $j=1$ for the first group, $i=1,\ldots,q_{1}$, and $j=2$
for the second group, $i=q_{1}+1,\ldots,q$.
Now, consider to expand the wavefunction $\phi_{i}$ 
by a complete orthonormal
basis set $\{\pi_l(r),(l=1,...,N)\}$ over $r\in[0,a]$ such that
\[
 \phi_i(r)=\sum_{l=1}^N c_{i,l}\pi_l(r).
\]
Substituting this into Eq.~(\ref{eq:2ch}), and integrating over the 
interval $[0,a]$, we obtain the $M=q\times N$-dimensional 
eigen value problem,
\begin{equation}
\left[\tilde H-\frac{i}{2}B-E I_{M}\right]\vec c=0,
\label{eq:2ch-mat}
\end{equation}
where
\begin{eqnarray*}
\tilde H&=&
\left(
\begin{array}{cccc}
\tilde{H}^{(1)} & U^{(12)} & \cdots & U^{(1q)}\\
U^{(12)} & \tilde{H}^{(2)} & \cdots & U^{(2q)}\\
\vdots & \vdots & \ddots & \vdots\\
U^{(q1)} & U^{(q2)} &\cdots& \tilde{H}^{(q)} 
\end{array}
\right),
\end{eqnarray*}
\begin{eqnarray*}
B=
\bordermatrix{
&1&\cdots &q_1& q_1+1 &\cdots &q{\rm -th\ block}\cr
&k_1 L \cr
& &\ddots &&&0 \cr
&&&k_1 L & &  \cr
&&&&k_2 L &   \cr
&&0 & & &\ddots\cr
&&&  &  && k_2 L}
\end{eqnarray*}
\begin{eqnarray*}
\vec c&=&
\left(\begin{array}{c}
c_{1,1}\\
\vdots\\
c_{1,N}\\
\vdots\\
c_{q,1}\\
\vdots\\
c_{q,N}
\end{array}\right)
\end{eqnarray*}
and
\begin{eqnarray*}
\tilde{H}^{(n)}_{ij}&=&\frac{1}{2}\int_0^a
\frac{d\pi_i}{dr}\frac{d\pi_j}{dr}dr+\int_0^a
\pi_iV_{n}\pi_jdr\nonumber\\
U^{(mn)}_{ij}&=&\int_{0}^{a}\pi_{i}V_{mn}\pi_{j}dr\\
L_{ij}&=&\pi_{i}(a)\pi_{j}(a).
\label{eq:2eig}
\end{eqnarray*}
In Eq.~(\ref{eq:2ch-mat}), $I_{M}$ is an $M$-dimensional unit matrix.
The eigen system Eq.~(\ref{eq:2ch-mat}) involves a pair of eigenvalues,
$k_{1}$ and $k_{2}$,
which may be rewritten as a standard eigenvalue equation for a single
variable $u$ according to the following heuristic procedure. Let us note
that energy $E$ can be represented by
both $k_{1}$ and $k_{2}$, namely
\[
E=\frac{1}{2}k_{1}^2+v_{1}=\frac{1}{2}k_{2}^2+v_{2}.
\]
so that 
\begin{equation}
(k_{1}+k_{2})(k_{1}-k_{2})=4\Delta^2
\label{eq:k1andk2}
\end{equation}
where
\[
 \Delta=\sqrt{\frac{v_2-v_1}{2}}.
\]
(Here, we assume $v_2\ge v_1$ for simplicity.)
Since the product of linearly independent combinations of $k_{1}$ and 
$k_{2}$ becomes constant, we require $k_{1} \pm k_{2}$ to 
satisfy the following conditions,
\begin{eqnarray*}
k_{1}+k_{2}&=&2i\Delta u \\
k_{1}-k_{2}&=&-2i\Delta u^{-1}.
\end{eqnarray*}
Thus,
\begin{eqnarray*}
k_{1}&=&i\Delta\left(u-u^{-1}\right)\\
k_{2}&=&i\Delta\left(u+u^{-1}\right)
\end{eqnarray*}
and 
\[
 E=\overline{v}-\Delta^2\frac{1+u^4}{2u^2}
\]
with
\[
 \overline{v}=\frac{v_1+v_2}{2}.
\]
This procedure of replacing a pair of variables $k_{1}$ and $k_{2}$ by a
single variable $u$ is called uniformization.

\subsection{The Tolstikhin-Siegert equation}
The uniformization described above reduces Eq.~(\ref{eq:2ch-mat}) to
\begin{equation}
{\cal M}(u)\vec{c}=0 
\label{eq:TSE2ch}
\end{equation}
with
\begin{eqnarray}
{\cal M}(u)=I_{M}+uB^-+u^2A+u^3B^++u^4I_{M},
\label{eq:M}
\end{eqnarray}
where
\begin{eqnarray*}
A&=&\frac{2}{\Delta^2}\left(\begin{array}{cccc}
\tilde{H}^{(1)}-\overline{v}I_{M} &U^{(12)} &\cdots&U^{(1q)} \\
U^{(12)}&\tilde{H}^{(2)}-\overline{v}I_{M} &\cdots&U^{(2q)} \\
\vdots & \vdots &\ddots&\vdots\\
U^{(1q)}&U^{(2q)} & \cdots &\tilde{H}^{(q)}-\overline{v}I_{M}
\end{array}
\right),\\
\end{eqnarray*}
and
\begin{eqnarray*}
B^{\pm}=
\frac{1}{\Delta}
\bordermatrix{
&1&\cdots &q_1& q_1+1 &\cdots &q{\rm -th\ block}\cr
&\pm  L \cr
& &\ddots &&&0 \cr
&&&\pm L & &  \cr
&&&& L &   \cr
&&0 & & &\ddots\cr
&&&  &  &&  L}.
\end{eqnarray*}
By introducing a new vector
\[
\left(
\begin{array}{c}
\vec c \\
u\vec c \\
u^2\vec c \\
u^3\vec c
\end{array}
\right),
\]
the non-linear eigenvalue problem, Eq.~(\ref{eq:TSE2ch}),
is reduced to a linear one such that
\begin{equation}
\left(
\begin{array}{cccc}
0 & I_{M} & 0 & 0 \\
0 & 0 & I_{M} & 0 \\
0 & 0 & 0 & I_{M} \\
-I_{M} & -B^{-} & -A & -B^{+} 
\end{array}
\right)
\left(
\begin{array}{c}
\vec c \\
u\vec c \\
u^2\vec c \\
u^3\vec c
\end{array}
\right)
=u
\left(
\begin{array}{c}
\vec c \\
u\vec c \\
u^2\vec c \\
u^3\vec c
\end{array}
\right).
\label{eq:TSE-2ch-2}
\end{equation}
Furthermore, the above equation is symmetrizable as follows,
\begin{eqnarray}
\left(
\begin{array}{cccc}
0 & 0 & 0 & I_{M} \\
0 & 0 & I_{M} & B^- \\
0 & I_{M} & B^- & A \\
I_{M} & B^{-} & A & B^{+} 
\end{array}
\right)
\left(
\begin{array}{c}
\vec c \\
u\vec c \\
u^2\vec c \\
u^3\vec c
\end{array}
\right) \nonumber \\
=u
\left(
\begin{array}{cccc}
0 & 0 & I_{M} & 0 \\
0 & I_{M} & B^- & 0 \\
I_{M} & B^{-} & A & 0 \\
0 & 0 & 0 & -I_{M} 
\end{array}
\right)
\left(
\begin{array}{c}
\vec c \\
u\vec c \\
u^2\vec c \\
u^3\vec c
\end{array}
\right).
\label{eq:TSE-2ch-3}
\end{eqnarray}
Let us refer to Eqs.~(\ref{eq:TSE2ch}), (\ref{eq:TSE-2ch-2}), and (\ref{eq:TSE-2ch-3}) 
as the Tolstikhin-Siegert equations (TSEs).
\section{FIRST AND SECOND ORDER PERTURBATION THEORY}
\label{sect:PT}
\subsection{Derivation of Perturbation Formulas}
Let us formulate the perturbation theory as appropriate for the SPS
whose orthonormality relation is different from the
standard one.
Relegating the one-threshold case to the next subsection, we treat
the general two-threshold case.
We assume the perturbing potential energy $V^{'}(r)$ 
vanishes beyond $r=a$, {\it i.e.},
\[
V'(r)=
\left(\begin{array}{cccc}
V'_{11} &V'_{12} & \cdots &V'_{1q}\\
V'_{12} &V'_{22} & \cdots &V'_{2q}  \\
\vdots & \vdots & \ddots&\vdots\\
V'_{1q} &V'_{2q}& \cdots & V'_{qq}\\
\end{array}
\right)=0\qquad(r > a).
\]
The TSE for the $n$-th state 
including perturbing potential energy reads
\begin{eqnarray}
\left(I_{M}+u_nB^-+u^2_nA+2\lambda\frac{u^2_nU^\prime}{\Delta^2}
+u^3_nB^++u^4I_{M}\right)\vec{c}_n=0,
\label{eq:ptb}
\end{eqnarray}
where
\begin{eqnarray*}
U'&=&
\left(\begin{array}{cccc}
U'^{(11)} &U'^{(12)} & \cdots &U'^{(1q)}\\
U'^{(12)} &U'^{(22)} & \cdots &U'^{(2q)}  \\
\vdots & \vdots & \ddots&\vdots\\
U'^{(1q)} &U'^{(2q)}& \cdots & U'^{(qq)}\\
\end{array}
\right),\\
U_{ij}^{\prime(mn)}&=&\int_0^a\pi_iV'_{mn}\pi_j dr.
\end{eqnarray*}
Differentiating Eq.~(\ref{eq:ptb}) with respect to $\lambda$ and 
using the orthonormal relationship (see Eq.(44) in Ref.~\cite{sit-tol}),
\[
\vec c_{m}^T\left[
I_{M}+\frac{u_mu_n(B^--u_mu_nB^+)}{(u_m+u_n)(1-u^2_mu^2_m)}
\right]\vec c_{n}=\delta_{mn},
\]
we obtain the Hellmann-Feynman theorem (HFT) in the present context, namely,
\begin{eqnarray}
\vec{c}^T_nU^\prime\vec{c}_n =\Delta^2\frac{1-u^4_n}{u^3_n}\frac{du_n}{d\lambda}=
\frac{d}{d\lambda}\left(\overline{v}-\Delta^2
\frac{1+u^4_n}{2u^2_n}\right)=\frac{dE_n}{d\lambda}.
\label{eq:HFT}
\end{eqnarray}
Now, we consider the perturbation series of $u_n$ and $\vec{c}_n$ such that
\begin{eqnarray}
 u_n&=&u^{(0)}_n+\lambda u^{(1)}_n+\lambda^2u^{(2)}_n+\cdots,\label{eq:pex1}\\
 \vec{c}_n&=&\vec{c}^{(0)}_n+\lambda\vec{c}^{(1)}_n
+\lambda^2\vec{c}^{(2)}_n+\cdots,\label{eq:pex2}
\end{eqnarray}
where $u_n^{(0)}$ and $\vec{c}_n^{(0)}$  are the $n$-th solution to
the unperturbed equation, Eq.~(\ref{eq:TSE2ch}),
\begin{eqnarray*}
{\cal M}(u_n^{(0)}) \vec{c}_n^{(0)}=0.
\end{eqnarray*}
Substituting the perturbation series, Eqs.~(\ref{eq:pex1}) and(\ref{eq:pex2}),
into Eq.~(\ref{eq:HFT}) and then comparing each power of $\lambda$, we obtain
\begin{eqnarray}
\lambda^0&:&\Delta^2\frac{1-u^{(0)4}_n}{u^{(0)3}_n}u_{n}^{(1)}
=\vec{c}^{(0)T}_n U^\prime \vec{c}^{(0)}_n\label{eq:lambda0}\\
\lambda^1&:&\frac{\Delta^2}{2u^{(0)3}_n}\left[
2u_n^{(2)}(1-u^{(0)4}_{n})-
\frac{3u_n^{(1)2}+u^{(1)2}_nu^{(0)4}_n}{u^{(0)}_n}\right]\nonumber\\
&&=\vec{c}^{(0)T}_n U^\prime \vec{c}^{(1)}_n.\label{eq:lambda1}
\end{eqnarray}
Next, let us evaluate the expansion coefficients over the
unperturbed eigenstates.
To this end, we rewrite Eq.~(\ref{eq:ptb})
using Eq.~(\ref{eq:M}), namely, 
\[
 {\cal M}(u_n)\vec{c}_n=-\frac{2\lambda u^2_n}{\Delta^2} U^\prime\vec{c}_n
\]
so that
\begin{equation}
  \vec{c}_n=-\frac{2\lambda u^2_n {\cal M}^{-1}(u_n)}{\Delta^2} U^\prime\vec{c}_n.
\label{eq:c}
\end{equation}
The spectral representation of ${\cal M}^{-1}$is given by
\[
 {\cal M}^{-1}(u_n)=\sum_{l=1}^{4M}
\frac{u^{(0)}_l\vec{c}^{(0)}_l\vec{c}_l^{(0)T}}{2(1-u_{l}^{(0)4})(u^{(0)}_l-u_n)}.
\]
(See Eq. (59) in Ref.~\cite{sit-tol}.)
Using the relations
\begin{eqnarray}
\sum_{l=1}^{4M}\frac{u_l^{(0)p} \vec{c}_l^{(0)}\vec{c}_l^{(0)T}}{1-u_l^{(0)4}}=0
 \qquad(p=1,2),
\end{eqnarray}
(see Eqs.~(51) and (52) in Ref.~\cite{sit-tol}), we have
\begin{eqnarray}
u_{n}^2 {\cal M}^{-1}(u_{n})=
\sum_{l=1}^{4M}
\frac{u^{(0)3}_l\vec{c}^{(0)}_l\vec{c}_l^{(0)T}}{2(1-u^{(0)4}_{l})(u^{(0)}_l-u_n)}.
\end{eqnarray}
Substituting this into Eq.~(\ref{eq:c}) and
comparing both hand sides power by power for $\lambda$,
and then using Eqs.~(\ref{eq:lambda0}) and (\ref{eq:lambda1}), we have
\begin{widetext}
\begin{eqnarray}
 \lambda^{0}: \vec{c}^{(0)}_n&=&\frac{1}{\Delta^2}\frac{u^{(0)3}_n\vec{c}^{(0)}_n\vec{c}^{(0)T}_n}{(1-u^{(0)4}_n)u^{(1)}_n}U^\prime\vec{c}^{(0)}_n=\vec{c}^{(0)}_n\\
\lambda^1: \vec{c}^{(1)}_n&=&\frac{1}{\Delta^2}
\sum_{l\not=n}^{4M}\frac{u^{(0)3}_l W^{\prime}_{ln}}{(1-u^{(0)4}_l)(u^{(0)}_n-u^{(0)}_l)}\vec{c}^{(0)}_l
-\frac{u_{n}^{(0)4}+3}{2u_{n}^{(0)}(1-u_{n}^{(0)4})}u_{n}^{(1)}\vec c_{n}^{(0)}\nonumber\\
&=&\frac{1}{\Delta^2}
\sum_{l\not=n}^{4M}\frac{u^{(0)3}_l W^{\prime}_{ln}}{(1-u^{(0)4}_l)(u^{(0)}_n-u^{(0)}_l)}\vec{c}^{(0)}_l+
\frac{W^{\prime}_{nn}}{2}
\left(\frac{1}{k_{1n}^{(0)2}}+\frac{1}{k_{2n}^{(0)2}}
-\frac{1}{k_{1n}^{(0)}k_{2n}^{(0)}}\right)\vec c_{n}^{(0)}
\label{eq:c2}
\end{eqnarray}
\end{widetext}
where
\[
 W^\prime_{mn}=\vec{c}^{(0)T}_m U^\prime\vec{c}^{(0)}_n
\]
and, as before,
\[
 k^{(0)}_{1n}=i\Delta[u^{(0)}_{n}-(u^{(0)}_{n})^{-1}],\qquad
 k^{(0)}_{2n}=i\Delta[u^{(0)}_{n}-(u^{(0)}_{n})^{-1}].
\]
Let us note that for $\vec{c}^{(1)}_n$, there is a term on top of the summation, which
is made absent in a standard perturbation theory because the
normalization is unchanged in so far as this term is purely imaginary
under the standard orthogonality relation. This freedom is not warranted
in the present case.

Finally, we have the perturbation formulas for the two-threshold SPS,
\begin{widetext}
\begin{eqnarray}
E_n^{(1)}&=&\vec{c}^{(0)T}_n U^\prime \vec{c}^{(0)}_n\\
 E^{(2)}_n&=&\vec{c}^{(0)T}_n U^\prime\vec{c}^{(1)}_n
=\frac{1}{\Delta^2}
\sum_{l\not=n}^{4M}\frac{u^{(0)3}_l W ^{\prime 2}_{ln}}{(1-u^{(0)4}_l)(u^{(0)}_n-u^{(0)}_l)}\nonumber\\
&+&
\frac{W^{\prime 2}_{nn}}{2}
\left(\frac{1}{k_{1n}^{(0)2}}+\frac{1}{k_{2n}^{(0)2}}
-\frac{1}{k_{1n}^{(0)}k_{2n}^{(0)}}\right).
\end{eqnarray}
\end{widetext}

\subsection{One-threshold case as a degenerate limit}
\label{subsect:one-threshold}
It is important to clarify the relationship between one- and two-threshold cases.
In the following, we prove that perturbation formulas for the one-threshold case
are obtained when we implement a limit of $v_{2} \to v_{1}$. 
In this limit, the following scaling clarified in Ref.~\cite{sit-tol},
\begin{equation}
A \to \frac{1}{\Delta^2}\tilde A,\quad
B^{\pm} \to \frac{1}{\Delta}\tilde B^{\pm},\quad
u \to \frac{-\kappa}{\Delta}
\end{equation}
reduces the two-threshold TSE to a one-threshold one, namely,
\begin{equation}
(\tilde A+\kappa \tilde B+\kappa^2I_{M})\vec c=0
\label{eq:TSE1}
\end{equation}
where 
\begin{eqnarray*}
\tilde A&=&2\left(\begin{array}{cccc}
\tilde{H}^{(1)}-\overline{v}I_{M} &U^{(12)} &\cdots&U^{(1q)} \\
U^{(12)}&\tilde{H}^{(2)}-\overline{v}I_{M} &\cdots&U^{(2q)} \\
\vdots & \vdots &\ddots&\vdots\\
U^{(1q)}&U^{(2q)} & \cdots &\tilde{H}^{(q)}-\overline{v}I_{M}
\end{array}
\right),\\
\end{eqnarray*}
and
\begin{eqnarray*}
\tilde B=
\bordermatrix{
&1&\cdots  &q{\rm -th\ block}\cr
&- L&\cdots&0 \cr
&\vdots &\ddots&\vdots  \cr
&0&\cdots& -L    \cr}.
\end{eqnarray*}
and $\kappa =ik_{1}=ik_{2}$. This scaling corresponds to the solution 
$k_{1}=k_{2}$ in Eq.~(\ref{eq:k1andk2}) when $v_{1} \to v_{2}$.
Note that the solution $k_{1}=-k_{2}$ in Eq.~(\ref{eq:k1andk2}) is unphysical
since the degenerate threshold here means the equivalence of asymptotic
wavefunctions in this limit.

Thus, the scaling leads us to the perturbation formulas for 
the one-threshold case, namely
\begin{eqnarray*}
\vec c^{(1)}_{n}&=&\sum_{l\not=n}^{2M}
\frac{W^\prime_{ln}}{k^{(0)}_l(k^{(0)}_n-k^{(0)}_l)}\vec c^{(0)}_l+
\frac{W^\prime_{nn}}{2k^{(0)2}_n}\vec c^{(0)}_{n}\\
E^{(1)}_n&=&W^\prime_{nn}=\vec{c}^{(0)T}_nV^\prime\vec{c}^{(0)}_n\\
E^{(2)}_n&=&\sum_{l\not=n}^{2M}
\frac{W^{\prime 2}_{ln}}{k^{(0)}_l(k^{(0)}_n-k^{(0)}_l)}+
\frac{W^{\prime 2}_{nn}}{2k^{(0)2}_n}. 
\end{eqnarray*}
Note that the summation runs over 
the branch of $k_{1}=k_{2}$, that is only over a half of the full non-degenerate space.
These correspond to the SPS representation of More's formulas\cite{more1}.
Our expressions for the first-order eigenvector and for the
second-order eigenenergy are
different from his\cite{more2}. The origin of the
discrepancy has been traced to an algebraic error in
More's derivation of the first-order wavefunction.
(One necessary term is unfortunately dropped during his derivation.) 
As a result of this, an extra term is restored in either formula.
Here, one important difference from the standard perturbation theory is that
no Hermitian conjugates appear in these formulas. This might suggest at
first that there would remain phase ambiguity. However, any {\it ad hoc}
additive phase would instead mar the orthogonality relation, that is what
is the relative phase in the standard theory is fixed in the SPS
theory, thus leaving no ambiguity with the phase of eigen
functions. It is thus worthwhile to see the consistency of the
orthonormality relation and the Siegert boundary condition for the
particular case of $\vec c^{(1)}_{n}$. 
This verification is worked out in Appendix.

\subsection{A Model Problem}
\begin{figure}
\begin{center}
\includegraphics[height=7cm,width=7cm]{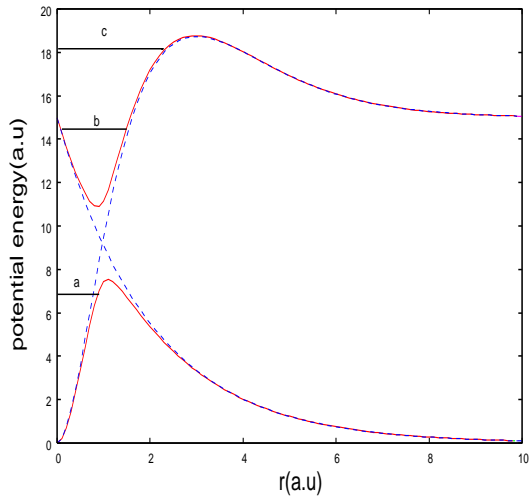}
\end{center}
\caption{Broken lines: Diagonal elements of the potential matrix
in Eq.~(\ref{eq:pot}). Solid lines: adiabatic potential energies.
This system supports three resonances:
~shape type~(a) in channel 1, Feshbach type~(b) and shape type~(c) in
 channel 2.} 
\label{fig:fig1}
\end{figure}

\begin{figure}
\begin{center}
\includegraphics[height=7cm,width=8cm]{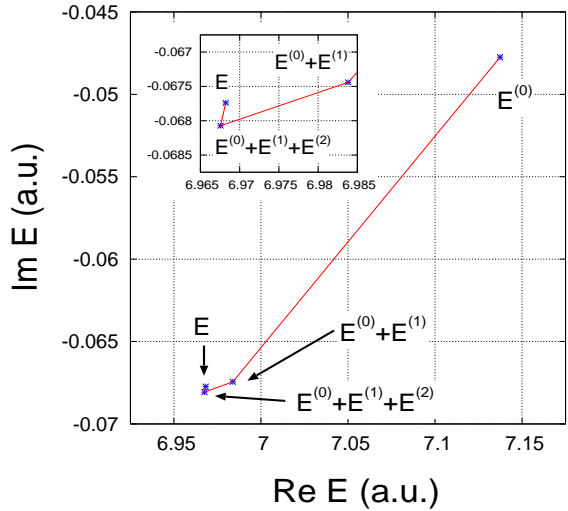}
\end{center}
\caption{Complex energies for resonance a}
\label{fig:fig2}
\end{figure}

\begin{figure}
\begin{center}
\includegraphics[height=7cm,width=8cm]{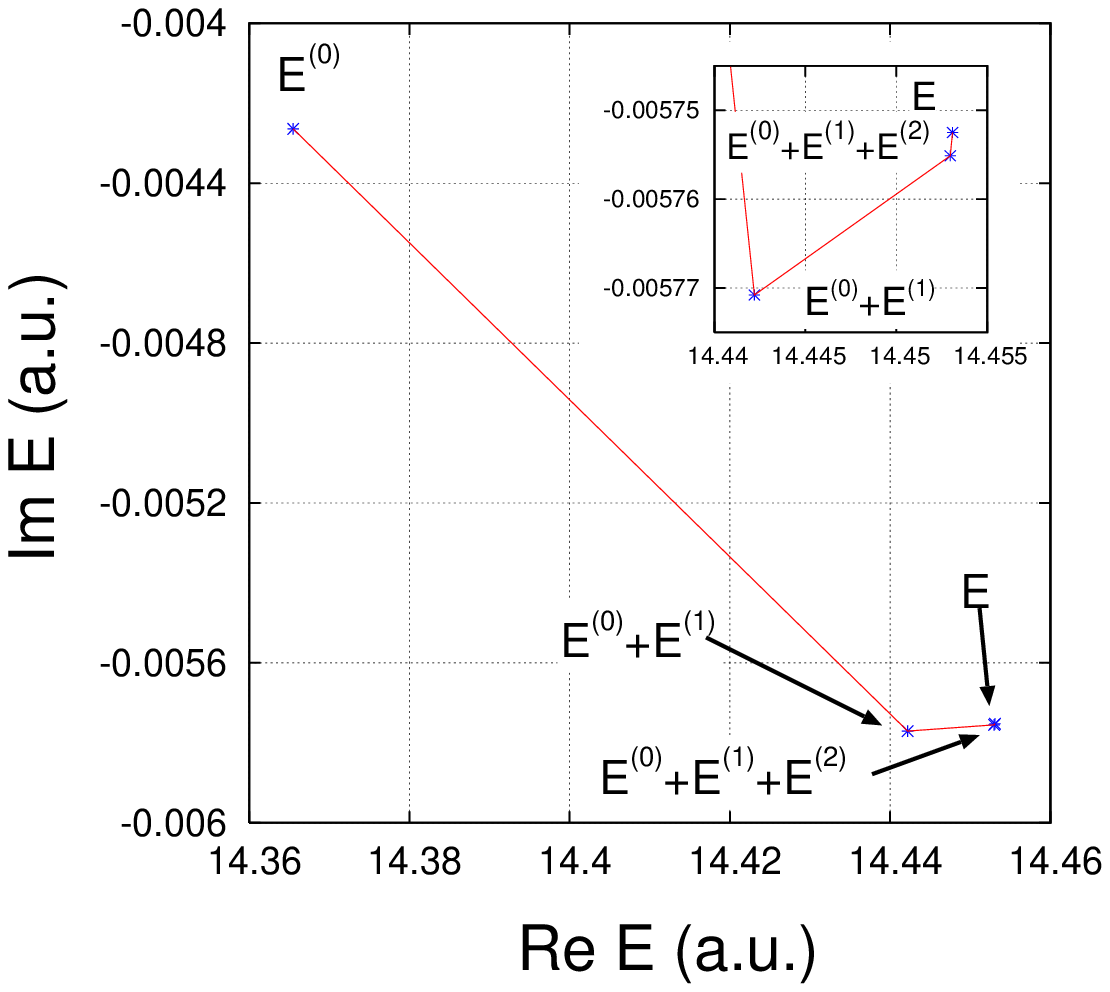}
\end{center}
\caption{Complex energies for resonance b}
\label{fig:fig3}
\end{figure}

\begin{figure}
\begin{center}
\includegraphics[height=7cm,width=8cm]{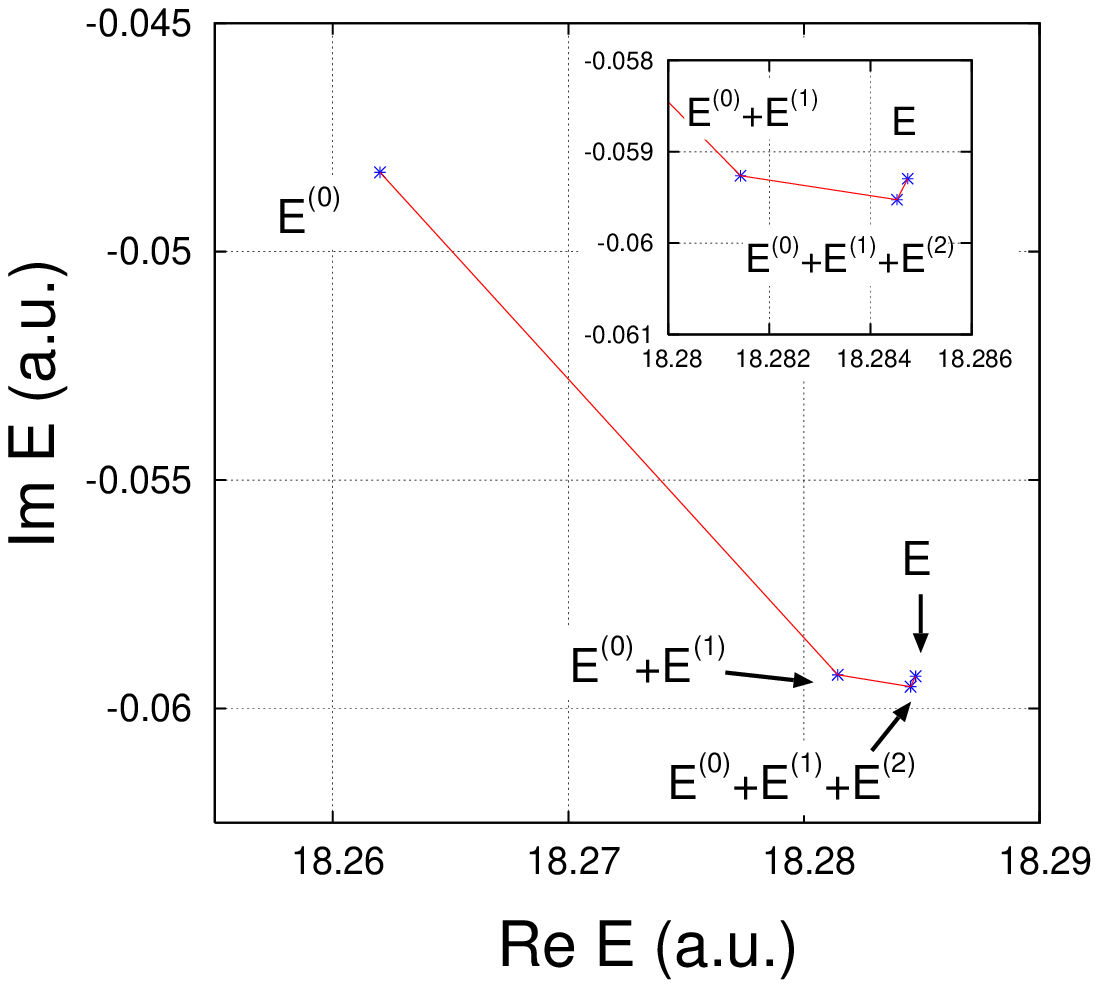}
\end{center}
\caption{Complex energies for resonance c}
\label{fig:fig4}
\end{figure}

Let us present an example of the perturbation theory for the
two-threshold case. 
We revisit the 2-channel model potential with two thresholds that is
taken up in Ref.~\cite{sit-tol}, {\it i.e.}
\begin{equation}
V(r)=
\left(\begin{array}{cc}
15e^{-0.5r} & 5re^{-r} \\
5re^{-r} & 15(r^2-r-1)e^{-r}+15
\end{array}
\right).
\label{eq:pot}
\end{equation}
The potential $V(r)$ supports three resonances. 
The adiabatic potential energy curve of the first channel supports
one shape type resonance~(a) while the other channel supports
one Feshbach type~(b) and one shape type~(c) resonance.
These resonances are depicted in Fig.~\ref{fig:fig1}.
We carried out the diagonalization of the TSE, Eq.~(\ref{eq:TSE-2ch-2}),
using the discrete variable representation (DVR) functions as a basis set.
The calculated resonance energies and widths with different
numbers of the basis functions are given in Table~\ref{table1}.
Let us call these results as {\it direct numerical solutions}.
To implement perturbation calculations, we separate $V(r)$ into
\[
 V(r)=V_{0}(r)+V^{'}(r)
\]
where
\begin{eqnarray*}
V_{0}(r)&=&\left(
\begin{array}{cc}
15e^{-0.5r} & 4re^{-r} \\
4re^{-r} & 15(r^2-r-1)e^{-r}+15
\end{array}
\right), \nonumber \\
V^{'}(r)&=&\left(
\begin{array}{cc}
0 & re^{-r} \\
re^{-r} & 0
\end{array}
\right). \\
\end{eqnarray*}
We regard $V_{0}$ as the unperturbed potential energy and $V^{'}$ 
as the perturbation potential energy.
We calculate perturbation energies using the
unperturbed solutions of TSE for
the same box size $a=50$ as in Ref.~\cite{sit-tol}.
Table \ref{table1} shows the results of first- and second-order 
perturbation calculations, and Figs.~\ref{fig:fig2}-\ref{fig:fig4} 
depict how the numerical solutions converge in the complex plane.
In the present model problem,
the first-order resonance energy
agrees with the direct numerical solutions to about 2 to 4 digits
while the width agrees to about 2 to 3 digits. 
And the second-oder resonance energy agrees to about 3 to 5 digits
while the width agrees to about 1 to 3 digits.
An important fact which we must remark is 
that the resonance energy and width do not appear to converge in pace.
For instance, the width of resonance ``c'' evaluated by 
the second-order perturbation theory appears
less accurate than the first-order one while the resonance energy appears
to have improved.
The seeming deterioration of the width is a little overwhelming,
all the more so for the improvement of the resonance energy.
Nonetheless, the distance between the second-order result and the direct
numerical one becomes rather small (see Fig.~\ref{fig:fig4}) in the
complex plane, that is in the Siegert state perturbation theory the
convergence is to be measured with respect to the variance
\begin{equation}
\label{eq:norm}
D= ||E-\sum_{n}\lambda^n E^{(n)}||
\end{equation}
rather than with respect to the real and imaginary parts of the sum,
individually. 

\begin{widetext}
\begin{table*}
\caption{Columns $\Re$,.
$\Im$, $D$, and $N$represent the real and imaginary parts of resonance
 energies, error variance in the complex plane, and the dimension of
 the DVR basis set, respectively.}
\label{table1}
\begin{tabular}{c ccc ccc ccc ccc}
\hline
\hline
&&\multicolumn{3}{c}{Resonance a}&&\multicolumn{3}{c}{Resonance b}
&&\multicolumn{3}{c}{Resonance c}\\
\cline{3-5}
\cline{7-9}
\cline{11-13}
$N$ &&  $\Re$ & $\Im$ & $D$ && $\Re$ & $\Im$ &  $D$ && $\Re$ & $\Im$ &  $D$ \\
\cline{1-1}
\cline{3-5}
\cline{7-9}
\cline{11-13}
&&\multicolumn{11}{c}{$E^{(0)}$}\\
  100 &   &    7.13731291  &   $-$0.04777819  &    0.17022398 &   &   14.36514823  &   $-$0.00441589  &    0.08759123 &   &   18.25940438  &   $-$0.04709964  &    0.02402711  \\
  300 &   &    7.13739307  &   $-$0.04774929  &    0.17034758 &   &   14.36548638  &   $-$0.00426431  &    0.08762023 &   &   18.26200618  &   $-$0.04826379  &    0.02526594  \\
  500 &   &    7.13739307  &   $-$0.04774929  &    0.17034758 &   &   14.36548638  &   $-$0.00426431  &    0.08762023 &   &   18.26200618  &   $-$0.04826379  &    0.02526594  \\
  700 &   &    7.13739307  &   $-$0.04774929  &    0.17034758 &   &   14.36548638  &   $-$0.00426431  &    0.08762023 &   &   18.26200619  &   $-$0.04826379  &    0.02526594  \\
   \\ 
&&\multicolumn{11}{c}{$E^{(0)}+E^{(1)}$}\\
  100 &   &    6.98368137  &   $-$0.06730063  &    0.01543038 &   &   14.44177638  &   $-$0.00607880  &    0.01094681 &   &   18.27770236  &   $-$0.05762106  &    0.00327772  \\
  300 &   &    6.98382603  &   $-$0.06744164  &    0.01560641 &   &   14.44219720  &   $-$0.00577079  &    0.01089678 &   &   18.28142974  &   $-$0.05926301  &    0.00330703  \\
  500 &   &    6.98382602  &   $-$0.06744165  &    0.01560641 &   &   14.44219720  &   $-$0.00577079  &    0.01089678 &   &   18.28142974  &   $-$0.05926301  &    0.00330703  \\
  700 &   &    6.98382603  &   $-$0.06744162  &    0.01560643 &   &   14.44219720  &   $-$0.00577079  &    0.01089678 &   &   18.28142974  &   $-$0.05926301  &    0.00330702  \\
   \\ 
&&\multicolumn{11}{c}{$E^{(0)}+E^{(1)}+E^{(2)}$}\\
  100 &   &    6.96760487  &   $-$0.06783608  &    0.00067084 &   &   14.45258871  &   $-$0.00611007  &    0.00013451 &   &   18.28074487  &   $-$0.05788235  &    0.00031413  \\
  300 &   &    6.96755505  &   $-$0.06807440  &    0.00074684 &   &   14.45297412  &   $-$0.00575530  &    0.00011988 &   &   18.28452186  &   $-$0.05952953  &    0.00031773  \\
  500 &   &    6.96755905  &   $-$0.06807406  &    0.00074312 &   &   14.45297384  &   $-$0.00575631  &    0.00012019 &   &   18.28452397  &   $-$0.05952538  &    0.00031324  \\
  700 &   &    6.96756395  &   $-$0.06807313  &    0.00073832 &   &   14.45297366  &   $-$0.00575513  &    0.00012033 &   &   18.28452412  &   $-$0.05952393  &    0.00031208  \\
   \\ 
&&\multicolumn{11}{c}{$E$ (Direct numerical solution)}\\
  100 &   &    6.96825547  &   $-$0.06767254 &              &   &   14.45272315  &   $-$0.00610584 &              &   &   18.28097965  &   $-$0.05767365 &               \\
  300 &   &    6.96822245  &   $-$0.06773922 &              &   &   14.45309397  &   $-$0.00575250 &              &   &   18.28473661  &   $-$0.05929537 &               \\
  500 &   &    6.96822245  &   $-$0.06773921 &              &   &   14.45309397  &   $-$0.00575250 &              &   &   18.28473661  &   $-$0.05929537 &               \\
  700 &   &    6.96822244  &   $-$0.06773923 &              &   &   14.45309397  &   $-$0.00575250 &              &   &   18.28473661  &   $-$0.05929536 &               \\
\hline
\hline
\end{tabular}
\end{table*}
\end{widetext}

\section{CONCLUSIONS}
In this paper we formulated one- and two-threshold SPSPT. 
The unusual orthonormality relationship of the SPSs 
results in somewhat nontrivial additional terms in SPSPT, and also it
determines the phase of the perturbation wavefunction. 
In the degenerate threshold case, the one-threshold SPSPT formulas are
obtained by appropriate scaling, and we also obtained an up-to-date correction
to More's theory.
The numerical calculations show how the perturbation results converge.
The convergence is achieved in the sense of the variance, Eq.~\ref{eq:norm}, but
not the resonance energy and width independently.

It is of interest to speculate on possible uses of SPSPT.
One immediate application would be to the manipulation of
Siegert poles. The shadow poles located near the physical sheet may be
transformed to physical 
resonances by an appropriate perturbation. We leave issues such as
this for a future task.

\section{ACKNOWLEDGMENT}
We thank Dr.~Tolstikhin for useful discussions.
This work was supported in part by Grants-in-Aid for Scientific Research
No.~15540381 
from the Ministry of Education, Culture, Sports, Science and Technology,
Japan, and also in part by the 21st Century COE program on ``Innovation
in Coherent Optical Science.''

\appendix

\section{Consistency with orthonormality relationship and Siegert
 boundary condition in first order}
Here, we prove that the first-order wavefunction satisfies the orthonormality
relationship and the Siegert boundary condition consistently.
First of all, we expand 
\begin{equation}
\vec c_{n}^{T}\left(I_{N}+\frac{1}{\kappa_{n}+\kappa_{m}}B\right)\vec c_{m}=\delta_{mn},
\label{eq:ortho}
\end{equation} 
into perturbation series, and compare
 both sides power by power for $\lambda$. The first-order equation shows
\begin{eqnarray}
&&\vec c_{n}^{(0)T}
\left(I_{N}-\frac{i}{k_{n}^{(0)}+k_{m}^{(0)}}B\right)\vec c_{m}^{(1)}
+c_{n}^{(0)T}
\frac{i(k_{n}^{(1)}+k_{m}^{(1)})}{(k_{n}^{(0)}+k_{m}^{(0)})^2} 
B\vec c_{m}^{(0)} \nonumber \\
&&+\vec c_{n}^{(1)T}\left(I_{N}-\frac{i}{k_{n}^{(0)}+k_{m}^{(0)}}B\right)c_{m}^{(0)}
=0.
\label{eq:ortho-1st}
\end{eqnarray}
And each term of the above equation reduces to
\begin{eqnarray*}
({\rm 1st~term})&=&\sum_{l \neq m}^{2N} 
\frac{W^{'}_{lm}\vec c_{n}^{(0)T}\vec c_{l}^{(0)}}
{k_{l}^{(0)}(k_{n}^{(0)}+k_{m}^{(0)})} \\
&+&\frac{2W^{'}_{nm}}{(k_{n}^{(0)}+k_{m}^{(0)})(k_{m}^{(0)}-k_{n}^{(0)})} 
+\frac{W^{'}_{mm}}{2k_{m}^{(0)2}}\delta_{nm},
\end{eqnarray*}
\begin{eqnarray*}
({\rm 2nd~term})=\frac{W^{'}_{nn}/k_{n}^{(0)}+W^{'}_{mm}/k_{m}^{(0)}}
{k_{n}^{(0)}+k_{m}^{(0)}}(\vec c_{n}^{(0)T}\vec c_{m}^{(0)}-\delta_{mn})
\end{eqnarray*}
and
\begin{eqnarray*}
({\rm 3rd~term})&=&\sum_{l \neq n}^{2N}
\frac{W^{'}_{ln}\vec c_{l}^{(0)T}\vec c_{m}^{(0)}}
{k_{l}^{(0)}(k_{n}^{(0)}+k_{m}^{(0)})} \\
&-&\frac{2W^{'}_{nm}}{(k_{n}^{(0)}+k_{m}^{(0)})(k_{m}^{(0)}-k_{n}^{(0)})} 
+\frac{W^{'}_{nn}}{2k_{n}^{(0)2}}\delta_{nm}.
\end{eqnarray*}
Hence, the left side of (\ref{eq:ortho-1st}) reduces to
\begin{eqnarray*}
(\ref{eq:ortho-1st})&=&\sum_{l=1}^{2N}
\frac{W^{'}_{lm}\vec c_{n}^{(0)T}\vec c_{l}^{(0)}}
{k_{l}^{(0)}(k_{n}^{(0)}+k_{m}^{(0)})}
+\sum_{l=1}^{2N}
\frac{W^{'}_{ln}\vec c_{l}^{(0)T}\vec c_{m}^{(0)}}
{k_{l}^{(0)}(k_{n}^{(0)}+k_{m}^{(0)})} \\
&+&\frac{W^{'}_{mm}}{k_{m}^{(0)}}
\left(\frac{1}{2k_{m}^{(0)}}-\frac{1}{k_{n}^{(0)}+k_{m}^{(0)}}\right)\delta_{mn}\\
&+&\frac{W^{'}_{mm}}{k_{m}^{(0)}}
\left(\frac{1}{2k_{m}^{(0)}}-\frac{1}{k_{n}^{(0)}+k_{m}^{(0)}}\right)\delta_{mn} \\
&=&\frac{1}{k_{n}^{(0)}+k_{m}^{(0)}}
\left(\sum_{l=1}^{2N}
\frac{W^{'}_{lm}\vec c_{n}^{(0)T}\vec c_{l}^{(0)}}{k_{l}^{(0)}}
+\sum_{l=1}^{2N}\frac{W^{'}_{ln}\vec c_{l}^{(0)T}\vec c_{m}^{(0)}}{k_{l}^{(0)}}
\right).
\end{eqnarray*}
By using a SPS sum rule,
\[
\sum_{l=1}^{2N}\frac{1}{k_{l}^{(0)}}\vec c_{l}^{(0)}\vec c_{l}^{(0)T}=0,
\]
we obtain
\[
\sum_{l=1}^{2N}\frac{1}{k_{l}^{(0)}}W^{'}_{lm}\vec c_{n}^{(0)T}\vec c_{l}^{(0)}
=\vec c_{n}^{(0)T}\left(\sum_{l=1}^{2N}\frac{1}{k_{l}^{(0)}}c_{l}^{(0)}
\vec c_{l}^{(0)T}\right)U^{'}\vec c_{m}^{(0)}=0
\]
and
\begin{eqnarray*}
\sum_{l=1}^{2N}\frac{1}{k_{l}^{(0)}}W^{'}_{ln}\vec c_{l}^{(0)T}\vec c_{m}^{(0)}
&=&\sum_{l=1}^{2N}\frac{1}{k_{l}^{(0)}}W^{'}_{nl}\vec c_{l}^{(0)T}\vec c_{m}^{(0)} \\
&=&\vec c_{n}^{(0)T}U^{'}\left(
\sum_{l=1}^{2N}\frac{1}{k_{l}^{(0)}}
\vec c_{l}^{(0)}\vec c_{l}^{(0)T}\right)\vec c_{m}^{(0)}=0.
\end{eqnarray*}
Therefore, the first-order wavefunction is consistent with the
orthonormality relationship.

Next, let us consider the Siegert boundary condition. 
We expand the Siegert boundary condition
and compare both sides power by power for $\lambda$. The first-order equation shows
\[
\left.
\left(\frac{d}{dr}-ik_{n}^{(0)}\right)\phi_{n}^{(0)}-ik_{n}^{(1)}\phi_{n}^{(0)}
\right |_{r=a}=0.
\]
Then by using the coordinate representation of the SPS sum rule, namely
\[
\sum_{l=1}^{2N}\frac{1}{k_{l}^{(0)}}\phi_{l}^{(0)}(r)\phi_{l}^{(0)}(r^{'})=0,
\]
we get
{\small
\begin{eqnarray*}
\left.
\left(\frac{d}{dr}-ik_{n}^{(0)}\right)\phi_{n}^{(0)}\right |_{r=a}
&=&\sum_{l \neq n}^{2N}\frac{W^{'}_{ln}}{k_{l}^{(0)}(k_{n}^{(0)}-k_{l}^{(0)})}
\left(\frac{d\phi_{l}^{(0)}}{dr}\right)_{r=a} \\
&-&k_{n}^{(0)}\sum_{l \neq n}^{2N}
\frac{W^{'}_{ln}}{k_{l}^{(0)}(k_{n}^{(0)}-k_{l}^{(0)})}\phi_{l}^{(0)}(a) \\
&=&-i\sum_{l \neq n}^{2N}\frac{1}{k_{l}^{(0)}}W^{'}_{ln}\phi_{l}^{(0)}(a) \\
&=&-i\sum_{l=1}^{2N}\frac{1}{k_{l}^{(0)}}W^{'}_{ln}\phi_{l}^{(0)}(a)
+i\frac{W^{'}_{nn}}{k_{n}^{(0)}}\phi_{n}^{(0)}(a) \\
&=&-i\int_{0}^{a}\left(\sum_{l=1}^{2N}\frac{1}{k_{l}^{(0)}}
\phi_{l}^{(0)}(a)\phi_{l}^{(0)}
\right)U^{'}\phi_{n}^{(0)}dr \\
&+&ik_{n}^{(1)}\phi_{n}^{(0)}(a) \\
&=&ik_{n}^{(1)}\phi_{n}^{(0)}(a)
\end{eqnarray*}
}
Hence the first-order wavefunction is consistent with the Siegert boundary condition.


\begin{references}
\bibitem{fano-rau} See, for instance, Chapter 8 in U.~Fano and
 A.~R.~P.~Rau, ``Atomic Collisions and Spectra'' (Academic Press,
 1986, New York), and references therein.
\bibitem{Siegert} A.~J.~F.~Siegert, Phys.~Rev.~{\bf 56}, 750 (1939).
\bibitem{oleg}O.~I.~Tolstikhin, V.~N.~Ostrovsky, and
 H.~Nakamura, Phys~Rev~A{\bf 58}, 2077 (1998).
\bibitem{olegclmb} O.~I.~Tolstikhin, I.~Yu.~Tolstikhina,~and~C.~Namba,~Phys.~Rev.~A{\bf 60}, 4673 (1999).
\bibitem{time-dept} S.~Yoshida, S.~Watanabe, C.~O.~Reinhold, and
J.~Burgd\"orfer, Phys.~Rev.~A{\bf 60}, 1113 (1999). 
\bibitem{CHG} R.~Santra, J.~M.~Shainline, and C.~H.~Greene,
 Phys.~Rev.~A{\bf  71}, 032703 (2005).
\bibitem{tana-wata}S.~Tanabe, S.~Watanabe, N.~Sato,
 M.~Matsuzawa, S.~Yoshida, C.~Reinhold, and
 J.~Burgd\"orfer, Phys.~Rev.~A{\bf 63}, 052721 (2001).
\bibitem{sit-tol} G.~V.~Sitnikov and O.~I.~Tolstikhin,
 Phys.~Rev.~A{\bf 67}, 032714 (2003).
\bibitem{toyota} K.~Toyota~and~S.~Watanabe,~Phys.~Rev.~A{\bf  68},
 062504 (2003).
\bibitem{more1} R.~M.~More,~Phys.~Rev.~A{\bf 3}, 1217 (1971).
\bibitem{more2} R.~M.~More,~Phys.~Rev.~A{\bf 4}, 1782 (1971).
\end{references}
\end{document}